\documentclass{article}
\begin{document}
\title{ The difficult discrimination of Impulse Stimulated Raman Scattering 
redshift against Doppler redshift 
\author{J. Moret-Bailly 
\footnote{Laboratoire de physique, Universit\'e de Bourgogne, BP 47870, F-21078 
Dijon cedex, France.
email : Jacques.Moret-Bailly@u-bourgogne.fr
}}}
\maketitle

\medskip
Pacs 42.65.Dr Stimulated Raman scattering, 94.10.Gb Absorption and scattering of radiation, 98.54.Aj 
Quasars 

\medskip
\textbf{\textit{Abstract}}
The Impulsive Stimulated Raman Scattering (ISRS) is a parametric light-matter interaction which shifts the 
frequencies of two ultrashort laser light pulses by a non-quantified transfer of energy. As ISRS has no 
threshold, the laser pulses may be replaced by the pulses which constitute the ordinary incoherent light. 
This replacement has the expected qualitative effect on the time constants required to observe ISRS: 
nanosecond collisional time and Raman period. It has also a qualitative effect, the frequency shifts become 
independent on the intensity; thus we use a new name for this avatar of ISRS: ''Incoherent Light Coherent 
Raman Scattering'' (ILCRS). 

The coherence makes ILCRS very different from the ordinary Raman effect proposed as an alternative to 
Doppler effect in the past: ILCRS is a stronger light-matter interaction, it does not blur the images, nether 
the spectra; the beams which receive energy are in the thermal radiation.

The shifts of the spectra produced either by a Doppler effect, or by ILCRS are very similar. However 
ILCRS is subject to a dispersion which perturbs slightly the spectra. ILCRS is the key of a model of 
quasars which explains all observations, without any new matter or physical theory: no fast moving cloud, 
no dark matter, no variation of the fine structure constant, no invisible object.

The redshifts and the thermal radiation produced by ILCRS should not be neglected {\it a priori}.

\section{Introduction} 

Twenty years ago, the interpretation of hundreds of observed redshifts seemed so difficult \cite{Reboul} 
than many alternatives were searched, in particular an incoherent Raman scattering \cite{Pecker}, but 
these trials failed because they led to a blur of the images, of the spectra, or they supposed strange 
properties of matter; the dangerous use of the photon \cite{WLamb,WLamb1} prevented to consider the 
coherent Raman scattering, as it tried to dissuade Townes from discovering the maser \cite{Townes}.

In previous papers \cite{M1, M2, M4}, we have described a light-matter interaction, named now 
"Incoherent Light Coherent Raman Scattering" (ILCRS), which shifts the frequency of incoherent light by 
the interference of an exciting incoherent beam with the light scattered by interaction with some low 
pressure gases. A few molecules per cubic metre would produce the whole cosmological redshift; this 
order of magnitude shows that a part of the redshift observed in the spectra of some objects could be 
produced by ILCRS.

As ILCRS is not well known, we need to describe it first. But, to avoid the heavy ab initio method used in 
the previous papers, ISRS will be considered here as an avatar of a well known effect: 

The ''Impulsive Stimulated Raman Scattering'' (ISRS), observed using high power ultrashort laser pulses, 
shifts the frequencies; these shifts depend on the intensity of the laser pulses \cite{Y,R,Wi,D1,D2,C}. 
Increasing the length of the pulses by orders of magnitude has important quantitative effects. ISRS has no 
intensity threshold, but decreasing the intensity has the usual linearisation effect due to the zero point 
electromagnetic field: the frequency shift becomes independent on the intensity; the new name ILCRS is 
justified by this qualitative change. 

\section{Extension of ISRS to long pulses} 
	The replacement of the short, femtosecond laser pulses usually used to perform ISRS, by pulses 
similar to the relatively long, nanosecond pulses which make incoherent light is a change of the scale of time 
by a factor of the order of $10^5$. This change of the scale of time applies to the other time constants:

i) The space-coherence of the Raman scattering of a wide beam requires that all mono- or poly-atomic 
molecules radiate with the same phase, that is that no collision perturbs the molecules during a pulse. In 
ISRS, this condition is fulfilled even in dense matter. To obtain ILCRS, a very low pressure gas must be 
used.

ii) The Raman period which corresponds to a transition between the molecular levels must be longer than 
the duration of the pulses: to be active, the gas must have Raman transitions in the radio-frequencies 
domain, that is it must have an hyperfine structure. Thus the active molecules are:

 - polyatomic molecules with an odd number of electrons; with some exceptions (NO), these molecules 
are not of common use in the labs even if they are stable, because they are chemically reactive, 
moderately, as OH, NH$_2$, or strongly as H$_2^+$; 

- heavy molecules or atoms;

- all mono- or poly-atomic molecules in an electric or magnetic field, because their eigenstates are split by 
Stark or Zeeman effects.

\section{Extension of Impulsive Stimulated Raman scattering (ISRS) to low power exciting 
light}

In previous papers, we considered the light emitted by dipolar molecules excited by an incident field, 
without collisions, in a spontaneous but coherent emission; the scattered electric field is proportional to the 
incident electric field, so that the frequency shift does not depend on the intensity of the incident field. In 
ISRS experiments \cite{Y}, the stimulated emitted field is proportional to the square of the incident field, 
so that it seems that the effects are different. They are not because the physically meaningful electric field 
$\hat E$ is the sum of the regular field $E$ obtained by the usual computation of the radiation of a dipole 
and the component $E_0$ of the zero point electromagnetic field which induces the emission of $E$; the 
field considered in ISRS, in a computation by the correct semiclassical electrodynamics or by quantum 
electrodynamics, is $\hat E = E_0+E$. Evidently, using lasers, $E_0$ is negligible. For a low level 
excitation, the zero point field is larger than $E$; in $\hat E^2=E_0^2+2EE_0+E^2$, $E^2$ is negligible 
and $E_0^2$ is a constant which exists in the dark, so that the square $\hat E^2$ varies proportionally to 
$E$ and the frequency shift does not depend on  the intensity of the field.

The common properties of ISRS and ILCRS are:

i) ISRS and ILCRS are space-coherent: The wave surfaces of an output beam are identical to the wave 
surfaces of the input beam; if the beam is wide enough to neglect diffraction, images brought by the beam 
remain good, without any blur.

ii) The input beam is frequency-shifted, but the width of a spectral line is not changed.

iii) The matter must have Raman lines; the period corresponding to a transition between the Raman levels 
must be longer than the duration of the pulses.

iv) The relative frequency-shift $\Delta\nu/\nu$ depends on the tensor of polarisability of the matter, which 
is subject to dispersion;

v) There is no intensity threshold.
\medskip

ILCRS differs from ISRS by:

vi) In ILCRS the frequency shift does not depend on the intensity of the beam. Thus $\Delta\nu/\nu$ is 
nearly constant.

\medskip
The shifts produced by ILCRS and Doppler effect are added; an evaluation of the ratio of the two shifts 
requires a study of the shifted spectrum precise enough to show the dispersion of ILCRS frequency shift.

\section{De-excitation of the molecules} 

The names ISRS and ILCRS are ambiguous because they mean that these effects are Raman, that is that a 
single light beam interacts with the molecules which are excited or de-excited. This Raman effect exists, but 
it is less intense than a four photons parametric effect which mixes two simultaneous Raman effects, so that 
the molecules keep their excitation constant. In laser experiments, the de-exciting beam may be produced 
by a laser, or result from a super-radiant generation. In ILCRS, the de-excitation of the very low pressure 
gas can only be radiative; we may have spontaneous radiative de-excitations, but more probably the 
parametric process, in which the ILCRS blueshift concerns mostly the 2.7K radiation, amplifying it.

The modes of the light have a temperature computed from Planck's laws. Transferring energy from a hot 
beam to a colder beam, the active molecules help an increase of the entropy of the light, they are a sort of 
catalyst.

\section{Observation of absorption lines of ILCRS active gases.}

Suppose that where a gas absorbs, the observed light beam is redshifted by ILCRS or the expansion of 
the Universe. The absorption scans the frequencies of the beam between the redshifted observed 
frequency and the line frequency, that is the width of the observed line nearly equals the redshift. As it is 
much widened, the line appears very weak; if there are many lines in the spectrum, the lines are mixed, the 
absorption appears uniform, the gas cannot be detected.

A consequence could be the non-detection of the molecule H$_2^+$ which is active in ILCRS. The 
observation of the 21 cm line shows that there are clouds of H$_2$; some astrophysicists think that 
H$_2$, only observable in absorption by this forbidden transition, could make a part of the dark matter. 
UV radiation is able to ionise H$_2$ into H$_2^+$, a very stable molecule; but H$_2^+$ is very 
reactive, destroyed by all collisions, so that while it may be relatively abundant in very low pressure 
H$_2$, on the contrary it is nearly absent at higher pressures. Thus, if H$_2^+$ is present in a cloud, the 
pressure is low enough for ILCRS : in any case it is not seen. On the contrary OH, NH$_2$,… are slowly 
destroyed by the collisions, they can exist at pressures which forbid ILCRS, so that they are observed.

In conclusion, invisible clouds of H$_2$ and H$_2^+$ can redshift the light which crosses them.

\section{Observation of the dispersion of ISRS}
The relative frequency shift $\Delta\nu/\nu$ by Doppler effect or expansion is strictly independent on the 
frequency. On the contrary ILCRS depends on the intensity of Raman scatterings subject to dispersion. 
Two reasons make this dispersion low : The first one, whose evaluation is very difficult, is that, in the 
parametric process, the dispersion of the redshifting Raman component is partly compensated by the 
influence of the blueshifting component. The second one comes from a compensation of the dispersions of 
two lines :

Write the ILCRS frequency shift in a sheet of a chemically homogenous, low pressure gas:
\begin{equation}
{\rm d} \nu = \frac{\rho\nu}{q+r(\nu)}{\rm d} x
\end{equation} where $\rho$ is the density of the gas and $q+r(\nu)$ a function depending on the gas in 
which $q$ is a constant, and the average of the small function $r(\nu)$ which represents the dispersion is 
zero.

Two frequencies $\nu_i$ and $\nu_j$ of the light source at x=0 and the corresponding frequencies 
$\nu_{io}$ and $\nu_{jo}$ observed at $x=X$ are bound to the mass of the gas per unit of surface by : 

\begin{equation}
\int_0^X\rho{\rm d} x=\int_{\nu_i}^{\nu_{io}}(q+r(\nu))\frac{{\rm d} \nu}{\nu}    
=\int_{\nu_j}^{\nu_{jo}}(q+r(\nu))\frac{{\rm d} 
\nu}{\nu}=\dots\label{dispersion}
\end{equation} 
Integrating :
\begin{equation} q\bigl[\ln\bigl(\frac{\nu_i}{\nu_{io}}\bigr)- 
\ln\bigl(\frac{\nu_j}{\nu_{jo}}\bigr)\bigr]=\int_{\nu_i}^{\nu_{io}} r(\nu)\frac{{\rm d} \nu}{\nu}-
\int_{\nu_j}^{\nu_{jo}} r(\nu)\frac{{\rm d} \nu}{\nu}= \int_{\nu_i}^{\nu_j} r(\nu)\frac{{\rm d} 
\nu}{\nu}-
\int_{\nu_{io}}^{\nu_{jo}} r(\nu)\frac{{\rm d} \nu}{\nu}\label{disp2}
\end{equation} 
Supposing that $\nu_i-\nu_j$ is small, $r(\nu)$ may be replaced by its mean value $m(\nu_i,\nu_j)$ 
between $\nu_i$ and $\nu_j$, so that equation \ref{disp2} becomes:
\begin{equation} \ln\bigl(\frac{\nu_i}{\nu_{io}}\bigr)- \ln\bigl(\frac{\nu_j}{\nu_{jo}}\bigr) \approx 
\frac{2(\nu_j-\nu_i) m(\nu_i,\nu_j)}{q(\nu_i+\nu_j)} -\frac{2(\nu_{jo}-\nu_{io}) 
m(\nu_{io},\nu_{jo})}{q(\nu_{io}+\nu_{jo})}\label{disp3}
\end{equation} Supposing that a frequency shift is purely Doppler, the first member of equation 
\ref{disp3} is zero; else, as all frequencies are known, equation \ref{disp3} gives a relation between the 
mean dispersions $ m(\nu_i,\nu_j)/q$ and $ m(\nu_{io},\nu_{jo})/q$.

Discriminating the dispersions in the intervals $(\nu_i,\nu_j)$ or $ (\nu_{io},\nu_{jo})$, seems possible by 
a statistical study if the lines are observed with many redshifts; the dispersions could characterise the 
redshifting gas. The spectroscopy of quasars  seems particularly favourable to such a characterisation 
because:

- observing similar quasars, sharp absorption lines are observed with many different redshift: 

- the active gases are made of atoms, so that there are few absorption lines, thus relatively large 
dispersions bound to resonances do not overlap;

-these resonances are near known absorption lines, so that their observation will be a test of the 
appropriateness of the theory for the quasars. Unhappily, up to now, a sufficient number of very high 
resolution, well calibrated spectra, is not available.

\section{The Lyman spectra of the quasars}
The spectrum of the quasar shows mainly Lyman lines of the hydrogen atom and some weaker UV metal 
lines. The strong emission spectrum defines the redshift of the quasar, while similar absorption spectra 
appear with several lower redshifted. The absorption lines are sharp.
Two types of theories have been proposed: either the hot atomic gas which absorbs is in a halo around the 
quasar, the redshift being produced by a Doppler effect, or it is in the intergalactic space, the redshift being 
produced by the expansion of the universe.

Supposing that the gas is next to the quasar, it must be ejected by the quasar with extremely large speeds to 
provide large enough redshifts; this requires  extremely large energies, and a propagation of the gas in the 
vacuum, so that a second problem is the confinement of the sheets of gas. As these problems cannot be 
solved reasonably, these types of theory are given up.

If the absorbing gas is in the intergalactic space, explanations of observations, maybe debatable, must be 
found:

a) The composition of the absorbing gas does not change much with the redshift

b)  There is a relation between density and temperature \cite{Hui}.

c) As the absorption lines are sharp, the absorbing clouds are thin in comparison with their distance; the 
diffusion of the clouds must be prevented by invisible matter.

d) It seems that the redshifts verify relations such as $\Delta {\rm ln}(z)=0,206$. A larger abundance of the 
clouds in old times is possible, but it is difficult to explain a precise rule of repartition of the clouds.

e) To remain over 10000K, the clouds must absorb from heating sources which do not appear clearly along 
the light paths.

f) According to Halton Arp \cite{Arp} many quasars are next to galaxies which have much lower redshifts

g) The relative frequency shift $\Delta\nu/\nu$ is not exactly constant in the spectrum. Webb et 
al.\cite{Wb} must suppose that the fine structure constant is not constant.

\medskip
Supposing that the atmosphere of the quasar observed by its emission spectrum extends into a static halo 
over several diameters of the quasar, and that a time-constant magnetic field varies along the light path, 
ILCRS explains immediately the sure and debatable observations:

a) As the halo is fed by the atmosphere it has a nearly constant composition;

b)	The density of the halo, and its heating decrease with the distance

c) The magnetic field induces Zeeman hyperfine structures in the gases which become active for ILCRS: as 
shown in previous section, while a light beam propagates in the halo, the frequencies of the light beam are 
decreased, so that the spectra written in the beam are redshifted; the eigen-frequencies of the gas, on the 
contrary, remain constant, so that, relative to the frequencies of the beam, these eigen-frequencies vary, 
writing absorption lines which are so wide and weak that they cannot be observed.

If the magnetic field is nearly zero, ILCRS does not appear and the lines are visibly written into the 
spectrum; several zeros of the magnetic field write a line with several redshifts. An elementary computation 
shows that the width at half intensity is nearly not increased, but that the lines have big feet \cite{M4}.

d) Many objects have a magnetic field; we may suppose, for instance, that an extremely low frequency 
evanescent electromagnetic field produced by the rotation of the core propagates in the halo; as the electrical 
conductivity of a plasma is large, the Stark effect may be neglected; the zeros of the magnetic field are 
evenly distributed, with our scale of time they appear stable. As the density of the halo decreases with the 
altitude, the variations of the redshifts decrease with the altitude too. The properties of the halo are nearly the 
same in all quasars, so that the law of distribution of the redshifts is nearly the same.

e) The core of the quasar heats the halo.

f) As ILCRS provides a large part of the redshift of the emission lines of the quasar, the quasar may be next to 
a galaxy having a lower redshift.

g)	As shown in the previous section, the dispersion induces a small variation of the ILCRS frequency 
shifts.

\section{Discussion}

ILCRS, an elementary optical effect, may explain a lot of observations with only the hypothesis of the 
existence of usual gases, eventually perturbed by electric or magnetic fields. The previous explanations 
have a single weakness, the absence of a reliable quantitative relation between the frequency shifts and the 
density of active gas; an evaluation of this relation done using a purely classical theory finds that the 
cosmological redshift could result of a constant density of 22 molecules per cubic metre. But a classical 
computation and the hypothesis are not reliable, so that this value may be wrong by orders of magnitude. 
The molecules which are active in ILCRS have a very rich spectra, so that a good determination of enough 
Raman tensors of polarisability is a terrible work.

An experimental measure of ILCRS requires a very low pressure, so that very long multipath cells are 
necessary; consequently, the used gas must be stable. NO may be used, but it is not an important molecule 
in space. However, we try to convince experimenters to set an ISRS experiment using nanosecond laser 
pulses, ILCRS parameters being easily deduced from ISRS ones if long pulses are used.

ILCRS is not a simple alternative to the Doppler effect; the energy lost by the redshift of hot beams is 
transferred to thermal radiation generally at 2.7K. Near the extremely bright objects, ILCRS or ISRS may 
be the source of the infrared radiation attributed to a hot dust which should be rejected by the pressure of 
radiation.

\section{Conclusion} 

The low pressure gases having a hyperfine spectrum are a sort of catalyst transferring energy from the hot 
beams of light to the cold ones or to the thermal radiation; this ''Coherent Raman Scattering of Incoherent 
Light'' produces a relative frequency shift of the hot beam, nearly constant in the spectrum. As clouds 
containing NO, NH$_2$\dots, or in magnetic fields are observed, a part of the redshifts is not produced 
by Doppler or expansion effects. Exceptionally well resolved and calibrated spectra of quasars show a non 
constant relative frequency shift attributed to a variation of the fine structure constant, more probably 
produced by ILCRS.

ILCRS gives an elementary explanation of probably all observations about the quasars with the simple 
hypothesis of a magnetic field in their halos. It seems to explain easily lots of other observations, for 
instance the thermal spectrum of bright objects without dust; it weakens the two main proofs of the big 
bang.

Will ILCRS solve many or only few problems? Whichever the answer, this powerful, elementary effect 
must be tested against the other explanations of non trivial observations of redshifts.

\end{document}